\documentclass[reprint, amsmath,amssymb, aps, ]{revtex4-2}

\usepackage[utf8]{inputenc} 
\usepackage{graphicx}
\usepackage{dcolumn}
\usepackage{bm}
\usepackage{amsmath}
\usepackage{enumitem}
\usepackage{hyperref} 
\usepackage[user]{zref}

\hypersetup{
	colorlinks=true,
	linkcolor=blue,
	citecolor=blue,
	filecolor=blue,
	urlcolor=blue,
}

\newcommand{\ve}[1]{\mathbf{#1}}
\newcommand{\te}[1]{\overline{\overline{#1}}}

\newcommand{\B}[1]{\textcolor[rgb]{0,0,0}{#1}}
\newcommand{\figref}[2]{\hyperref[#1]{Fig. \ref*{#1}{#2}}}

\newcommand{\eqqref}[1]{\hyperref[#1]{Eq.~\eqref{#1}}}

\def\@#1{_{\rm #1}}

\begin{document}

\preprint{APS/123-QED}

\title{Scattering Symmetries in Diffraction Gratings}

\author{Karim Achouri}%
\email{karim.achouri@epfl.ch}
\affiliation{%
Institute of Electrical and Microengineering, École Polytechnique Fédérale de Lausanne, Laboratory for Advanced Electromagnetics and Photonics, Lausanne, Switzerland\\}%

\date{\today}

\begin{abstract}

Metasurfaces enable powerful control of electromagnetic waves using subwavelength planar structures, but their deeply subwavelength periodicity typically suppresses propagating diffraction orders, which limits the number of available scattering channels. Diffraction gratings and metagratings overcome this limitation by supporting multiple propagating diffraction orders, thus providing additional degrees of freedom for controlling wave propagation. However, when several diffraction channels are present, it becomes nontrivial to predict how spatial symmetries combined with reciprocity affect the overall scattering response. For this purpose, we develop a formalism to determine the scattering symmetries of diffraction gratings supporting multiple diffraction orders. The approach is based on constructing a global scattering matrix that connects all incident and scattered diffraction channels and on introducing matrix representations of spatial symmetry operations acting on the field amplitudes. From these representations, we derive an invariance condition that directly constrains the sub-scattering matrices associated with each pair of diffraction orders. This provides a rigorous approach for computing the grating scattering coefficients imposed by symmetry and reciprocity. We illustrate the application of this approach via several examples and show how metagratings may be used to achieve, for instance, angle-asymmetric transmission and extrinsic chiral effects.

\end{abstract}


\maketitle

\section{Introduction}
\label{sec_intro}

In recent years, metasurfaces have emerged as a powerful platform for controlling electromagnetic waves using ultrathin planar structures composed of subwavelength scatterers. By engineering the local amplitude, phase, and polarization response of these scatterers, metasurfaces enable a wide variety of wavefront transformations, including anomalous reflection and refraction, beam steering, and flat optical components such as metalenses and holographic devices~\cite{yu2011,glybovski2016}. These capabilities have made metasurfaces an important tool for implementing compact photonic systems that realize complex electromagnetic functionalities.

A particularly important aspect of most metasurfaces is that their unit-cell periodicity is deeply subwavelength. As a consequence, only the zeroth diffraction order typically propagates, while higher diffraction orders remain evanescent. In many applications this property is advantageous because it suppresses parasitic diffraction and allows the metasurface to behave effectively as a homogenized interface characterized by local reflection and transmission coefficients~\cite{achouri2021}. However, this suppression of propagating diffraction orders also restricts the number of available scattering channels that can be independently controlled and thus limits the number of available degrees of freedom for wavefront shaping.

In contrast, diffraction gratings and, more recently, \emph{metagratings} exploit the presence of multiple propagating diffraction orders to manipulate the propagation of light. Metagratings consist of periodic arrangements of engineered scatterers designed to redistribute energy among a discrete set of diffraction orders~\cite{radi2017,popov2018,popov2019a,wang2024,rane2024}. Because each propagating diffraction order represents an \emph{independent scattering channel}, such structures provide additional degrees of freedom for tailoring the scattering response compared with conventional metasurfaces. This capability has enabled highly efficient beam steering, anomalous reflection, and diffraction-based wavefront control across a wide range of frequencies, from microwave to optical regimes~\cite{radi2017,popov2018,popov2019a,wang2024,rane2024}. Consequently, metagratings have, over the past few years, attracted a growing interest as an alternative and/or a complementary platform to metasurfaces for wavefront engineering.  

Despite these advances, predicting the scattering behavior of diffraction gratings supporting multiple diffraction orders remains challenging. In particular, the connection between the spatial symmetries of the structure and some fundamental physical constraints, such as reciprocity, can lead to nontrivial relationships between the scattering coefficients associated with different diffraction channels. When several diffraction orders are simultaneously present, these constraints can lead to complex and unintuitive symmetry relations in the grating global scattering response. Understanding how structural symmetries combine with reciprocity to determine the allowed scattering response of the system is therefore essential for both analyzing and designing devices that rely on diffraction control.

In this work, we develop a formalism to determine the scattering symmetries of diffraction gratings supporting multiple diffraction orders. Our approach is based on constructing a global scattering matrix that explicitly connects all incident and scattered diffraction channels together. Following the same approach as the one developed in~\cite{dmitriev2011,dmitriev2013a,achouri2023a}, we introduce matrix representations of spatial symmetry operations and derive an invariance condition that directly constrains the sub-scattering matrices associated with each pair of diffraction orders. This framework allows one to determine, in a rigorous and systematic fashion, which scattering coefficients must vanish and which must be equal due to spatial symmetries and reciprocity.

Using this formalism, we analyze several gratings exhibiting different spatial symmetries and illustrate how their diffraction responses are affected by symmetry and reciprocity. Additionally, we show how this framework can help to design diffraction-based devices, including structures exhibiting asymmetric transmission and extrinsic chiral effects.

\section{Theoretical formalism}
\label{sec_theo} 

Let us consider a periodic array lying in the $xy$-plane at $z=0$. It is illuminated by a TE or TM polarized plane wave incident within the $xz$-plane. The array has a rectangular lattice with periods $p_x$ and $p_y$ along the $x$- and $y$-directions, respectively. We consider that $p_y \ll \lambda$, where $\lambda$ is the wavelength of the waves in either of the $\pm z$ regions, so that \emph{the diffraction orders are restricted to the plane of incidence chosen to be the $xz$-plane}. Note that, even though all waves propagate within the $xz$-plane, it does not necessarily imply that the problem may be reduced to a two-dimensional analysis. This is because the scattering particles composing the array may be asymmetric along the $y$-direction. 

Following this convention, the \B{normalized} longitudinal components of the scattered diffraction orders wave vector, \B{$\hat{k}_{x,\text{s}}$ and $\hat{k}_{y,\text{s}}$}, either in reflection or transmission, are given by
\B{ 
\begin{equation}
	\label{eq_GEQ}
        \hat{k}_{x,\text{s}} = \hat{k}_{x,\text{i}} - d \mathcal{G} \quad \text{and} \quad \hat{k}_{y,\text{s}} = \hat{k}_{y,\text{i}} = 0,
\end{equation}
where
$\hat{k}_{x,\text{i}} = n_\text{i}\sin\theta_\text{i}$ is the normalized $x$-component of the incident wave wavevector, $\mathcal{G}=\lambda_0/p_x$ is the inverse of the normalized grating period with $\lambda_0$ being the freespace wavelength, $d\in \mathbb{N}$ is the diffraction order number and,  $n_\text{i}$ and $n_\text{s}$ are the refractive indices experienced by the incident and scattered waves (for reflected orders $n_\text{s}=n_\text{i}$), respectively.  
}
	
\begin{figure}[h!]
	\centering
	\includegraphics[width=\linewidth]{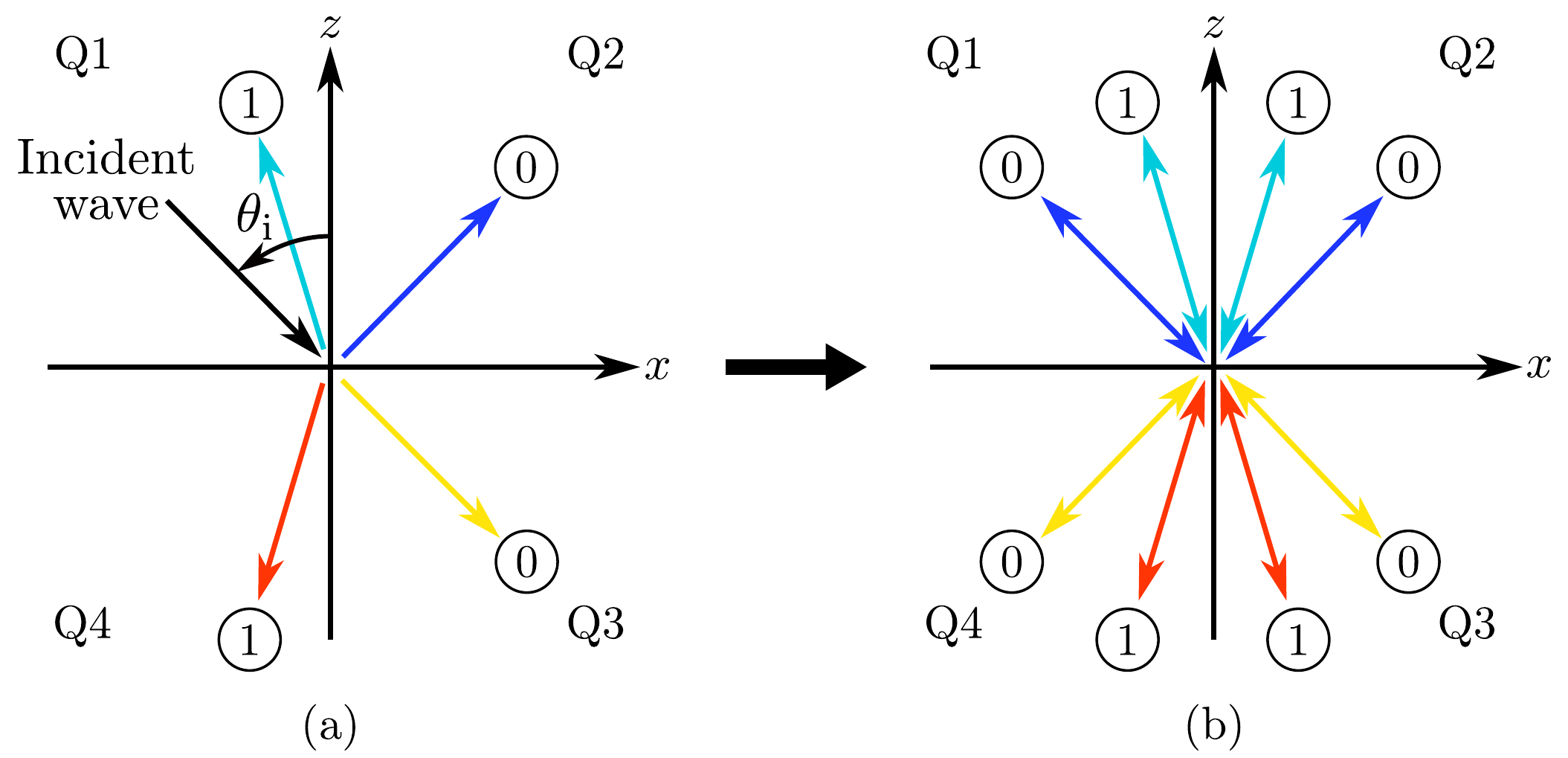}
        \caption{\B{From diffraction orders to the definition of open scattering channels. a) TM diffraction orders for a grating with parameters $n_\text{s}=n_\text{i}=1$, $\mathcal{G}=1$ and $\theta_\text{i}=45^\circ$. b) Convention used to define the open channels in all four quadrants. Since only TM waves are considered, only quadrants 1 to 4 are depicted.}}
	\label{fig_do2chan}
\end{figure}

\begin{figure}[h!]
	\centering
	\includegraphics[width=\linewidth]{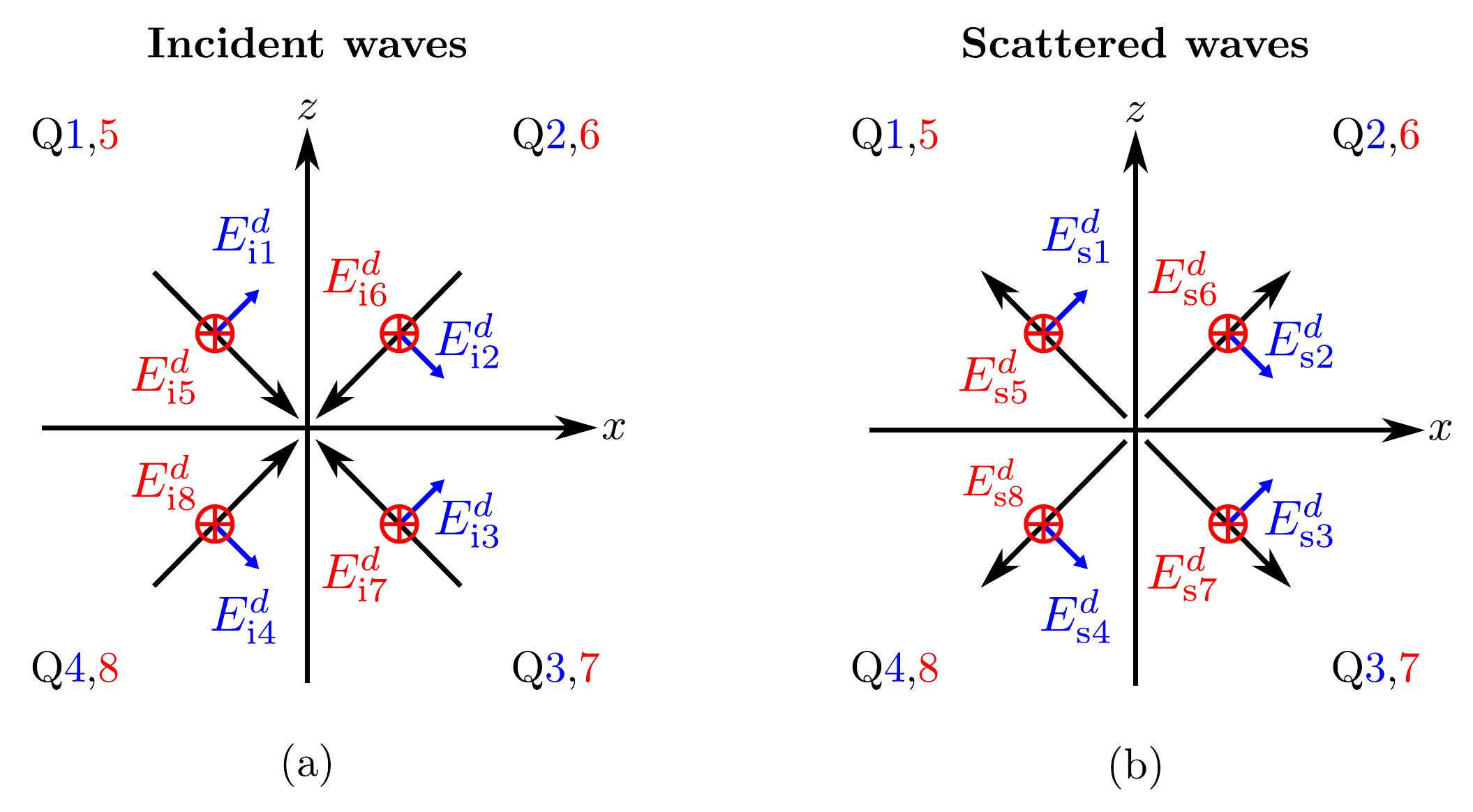}
	\caption{Naming and polarization conventions for the field amplitudes of the (a) incident and (b) scattered waves propagating along a given diffraction order channel $d$.}
	\label{fig_schematicdo}
\end{figure}

Now, our goal is to create a scattering matrix that connects together the incident and scattered waves interacting with a given diffraction grating. \B{This scattering matrix is constructed by first defining the properties of the array and its scattering channels from the parameter set $D = \{n_\text{i},n_\text{s},\theta_\text{i},\mathcal{G}\}$.} As an example, consider the diffraction scenario illustrated in Fig.~\ref{fig_do2chan}a for the specific set $D_1=\{1,1,45^\circ,1\}$. In our formalism, we have defined that each of the four quadrants of the $xz$-plane is labeled from Q1 to Q4 (for TM waves), starting from the top-left region and rotating clock-wise. \B{The same regions are labeled from Q5 to Q8 for TE waves, as shown in Fig.~\ref{fig_schematicdo}.} Using~\eqref{eq_GEQ} with $D_1$, the space propagating diffraction orders allowed to exist (open channels) are those for which $d = \{0,1\}$ in both $\pm z$ regions, as indicated by the numbers in the black circles in Fig.~\ref{fig_do2chan}a.

To build the scattering matrix, we now consider that each of these open channels is distributed symmetrically around the $z$-axis such that Q1 (Q5) and Q2 (Q6) share the same open channels and, similarly, Q3 (Q7) and Q4 (Q8) also share the same channels. Note that we separate the channels between the $\pm z$ regions since different refractive indices may be specified in these regions thus leading to different scattering angles in both regions. This is illustrated in Fig.~\ref{fig_do2chan}b where we further consider that each channel may serve as an input or an output to the system. With this formalism, the initially specified incident wave necessarily impinges on the grating via a $0^\text{th}$-order channel. Finally, the naming and polarization conventions used to define the complex amplitude of the incident and scattered fields are depicted in Fig.~\ref{fig_schematicdo}.

The general scattering matrix, $\te{S}$, may now be defined as
\begin{equation}
	\label{eq_GSM}
	\ve{E}_\text{s} = \te{S}\cdot\ve{E}_\text{i},
\end{equation}
where $\ve{E}_\text{i}$ and $\ve{E}_\text{s}$ are vectors of vectors containing the complex amplitudes of the incident and scattered fields, respectively, and are given by 
\begin{equation}
\label{eq_Ea}
\ve{E}_\text{a} = 
\left[
\ve{E}_\text{a}^{L}\,		\ve{E}_\text{a}^{L+1}\,		\cdots\,		\ve{E}_\text{a}^{U-1}\,		\ve{E}_\text{a}^{U}	\right]^T,
\end{equation}
where $a = \{\text{i},\text{s}\}$ which stands for the incident and scattered waves, respectively, and $L$ and $U$ are the lowest and highest values of $d$ (diffraction order number) in both the $\pm z$ regions, respectively. To compute the values of $L$ and $U$ in both $\pm z$ regions, we use
\B{\begin{subequations} 
\begin{align}
U_\text{a} &= \left\lfloor\left(n_\text{i}\sin\theta_\text{i}+n_\text{a}\right)\mathcal{G}^{-1}\right\rfloor, \\
L_\text{a} &= \left\lceil\left(n_\text{i}\sin\theta_\text{i}-n_\text{a}\right)\mathcal{G}^{-1}\right\rceil, 
\end{align} 
\end{subequations}}
where $a = \{\text{i},\text{s}\}$, $\lceil\cdot\rceil$ and $\lfloor\cdot\rfloor$ correspond to the ceiling and floor functions, respectively. It follows that
\begin{equation} 
    L = \min(L_\text{i},L_\text{s}) \quad\text{and}\quad U = \max(U_\text{i},U_\text{s}).
\end{equation}
And the total number of diffraction orders, $N_\text{a}$, in either regions is generally given by
\begin{equation} 
    \label{eq_Ntot} 
N_\text{a} = U_\text{a} - L_\text{a} + 1. 
\end{equation}
As an example, for the scenario depicted in Fig.~\ref{fig_do2chan}, we have that $L=0$ and $U=1$ with $N_\text{i} = N_\text{s} = 2$ since both regions have identical media. Note that this approach is used so as to always generate a matrix $\te{S}$ that is square. Naturally, this implies that if $L_\text{i}\neq L_\text{s}$ and/or $U_\text{i}\neq U_\text{s}$, some additional incident/scattered channels, that should in principle not be considered opened, are nonetheless artificially defined as opened. This issue is simply resolved by setting the corresponding problematic scattering coefficients in $\te{S}$ to zero.  

The vectors $\ve{E}_\text{a}^d$ in~\eqref{eq_Ea} contain the complex amplitude of the TM and TE incident and scattered waves in each of the quadrants (see Fig.~\ref{fig_schematicdo}) as
\begin{equation}
\ve{E}_\text{a}^d=[E_\text{a1}^d\,\, E_\text{a2}^d\,\, E_\text{a3}^d\,\, E_\text{a4}^d\,\, E_\text{a5}^d\,\, E_\text{a6}^d\,\, E_\text{a7}^d\,\, E_\text{a8}^d]^T.
\end{equation}
With this approach, we can express the general scattering matrix in~\eqref{eq_GSM} as
\begin{equation}
    \label{eq_genS} 
\te{S}=
\begin{bmatrix}
\te{S}^{LL} & \te{S}^{LL+1} & \cdots  &\te{S}^{LU-1} & \te{S}^{LU} \\
\te{S}^{L+1L} & \te{S}^{L+1L+1} & \cdots  &\te{S}^{L+1U-1} & \te{S}^{L+1U} \\
\vdots & \vdots & \ddots& \vdots & \vdots \\
\te{S}^{U-1L} & \te{S}^{U-1L+1} & \cdots  &\te{S}^{U-1U-1} & \te{S}^{U-1U} \\
\te{S}^{UL} & \te{S}^{UL+1} & \cdots  &\te{S}^{UU-1} & \te{S}^{UU} \\
\end{bmatrix}.
\end{equation}
This scattering matrix connects each input channel to each output channel as defined from the conventions in Fig.~\ref{fig_do2chan} and Fig.~\ref{fig_schematicdo}. The general scattering matrix in~\eqref{eq_genS} is composed of sub-scattering matrices that connect incident waves impinging on the grating from the diffraction channels of number $n$ and being scattering into the diffraction channels of number $m$. Concretely, this means that  
\begin{equation}
\label{eq_subS} 
\ve{E}_\text{s}^m = \te{S}^{mn}\cdot\ve{E}_\text{i}^n.
\end{equation}
From a general perspective, the shape of the sub-scattering matrix in~\eqref{eq_subS} takes two different forms. The first one corresponds to cases where the scattered waves exist in a quadrant longitudinally (along the $x$-direction) opposite to that of the considered incident wave, e.g., Q2 and Q3 for an incident wave coming from Q1. This is typically the case for scattering between channels of identical orders. Referring to Fig.~\ref{fig_do2chan}b, this would be the case for scattering events between channels $0\leftrightarrow 0$ and $1\leftrightarrow 1$. Since such scattering events preserve the sign of the scattered wave vector component, we shall refer to them as ``forward'' scattering. The corresponding forward sub-scattering matrix is given by
\begin{equation} 
\label{eq_FS} 
\te{S}^{mn}_\text{F}=
\begin{bmatrix}
0 & S_{12}^{mn} & S_{13}^{mn} & 0 & 0 & S_{16}^{mn} & S_{17}^{mn} & 0\\
S_{21}^{mn} & 0 & 0 & S_{24}^{mn} & S_{25}^{mn} & 0 & 0 & S_{28}^{mn}\\
S_{31}^{mn} & 0 & 0 & S_{34}^{mn} & S_{35}^{mn} & 0 & 0 & S_{38}^{mn}\\
0 & S_{42}^{mn} & S_{43}^{mn} & 0 & 0 & S_{46}^{mn} & S_{47}^{mn} & 0\\
0 & S_{52}^{mn} & S_{53}^{mn} & 0 & 0 & S_{56}^{mn} & S_{57}^{mn} & 0\\
S_{61}^{mn} & 0 & 0 & S_{64}^{mn} & S_{65}^{mn} & 0 & 0 & S_{68}^{mn}\\
S_{71}^{mn} & 0 & 0 & S_{74}^{mn} & S_{75}^{mn} & 0 & 0 & S_{78}^{mn}\\
0 & S_{82}^{mn} & S_{83}^{mn} & 0 & 0 & S_{86}^{mn} & S_{87}^{mn} & 0
\end{bmatrix}.
\end{equation}
In $\te{S}^{mn}_\text{F}$, the top-left $4\times4$ elements correspond to TM to TM scattering, the top-right elements correspond to TE to TM scattering, the bottom-left correspond to TM to TE scattering and the bottom-right correspond to TE to TE scattering. 

The second type of sub-scattering matrix corresponds to scattering events that do not preserve the sign of the scattered wave vector component. In such cases, the scattered waves propagate backward with respect to the direction of propagation of the considered incident wave, e.g., in Q1 and Q4 for an incident wave coming from Q1. Referring again to the example depicted in Fig.~\ref{fig_do2chan}, this would be the case for scattering events between channels $0\leftrightarrow1$. The corresponding backward sub-scattering matrix is defined as the complementary of $\te{S}^{mn}_\text{F}$ and is given by
\begin{equation} 
\label{eq_BS} 
\te{S}^{mn}_\text{B}=
\begin{bmatrix}
S_{11}^{mn} & 0 & 0 & S_{14}^{mn} & S_{15}^{mn} & 0 & 0 & S_{18}^{mn} \\
0 & S_{22}^{mn} & S_{23}^{mn} & 0 & 0 & S_{26}^{mn} & S_{27}^{mn} & 0 \\
0 & S_{32}^{mn} & S_{33}^{mn} & 0 & 0 & S_{36}^{mn} & S_{37}^{mn} & 0 \\
S_{41}^{mn} & 0 & 0 & S_{44}^{mn} & S_{45}^{mn} & 0 & 0 & S_{48}^{mn} \\
S_{51}^{mn} & 0 & 0 & S_{54}^{mn} & S_{55}^{mn} & 0 & 0 & S_{58}^{mn} \\
0 & S_{62}^{mn} & S_{63}^{mn} & 0 & 0 & S_{66}^{mn} & S_{67}^{mn} & 0 \\
0 & S_{72}^{mn} & S_{73}^{mn} & 0 & 0 & S_{76}^{mn} & S_{77}^{mn} & 0 \\
S_{81}^{mn} & 0 & 0 & S_{84}^{mn} & S_{85}^{mn} & 0 & 0 & S_{88}^{mn}
\end{bmatrix}.
\end{equation}

\B{We now use~\eqref{eq_FS} and~\eqref{eq_BS} to define the sub-scattering matrix in~\eqref{eq_subS}. To do so, we must consider the sign of $\hat{k}_{x,\text{s}}$ and determine whether it differs from that of the considered incident wave in~\eqref{eq_subS}. If they have the same sign, then $\te{S}^{mn} = \te{S}_\text{F}^{mn}$. If they have opposite sign, then $\te{S}^{mn} = \te{S}_\text{B}^{mn}$. Finally, if either of the considered incident wave or the corresponding scattered wave propgate normally with respect to the grating, then $\te{S}^{mn} =\te{S}_\text{F}^{mn} + \te{S}_\text{B}^{mn}$.}


It should be noted that the proposed formalism leads to redundant and problematic definitions of the diffraction order channels in the following two situations. 

\begin{figure}[h!]
\centering
\includegraphics[width=\linewidth]{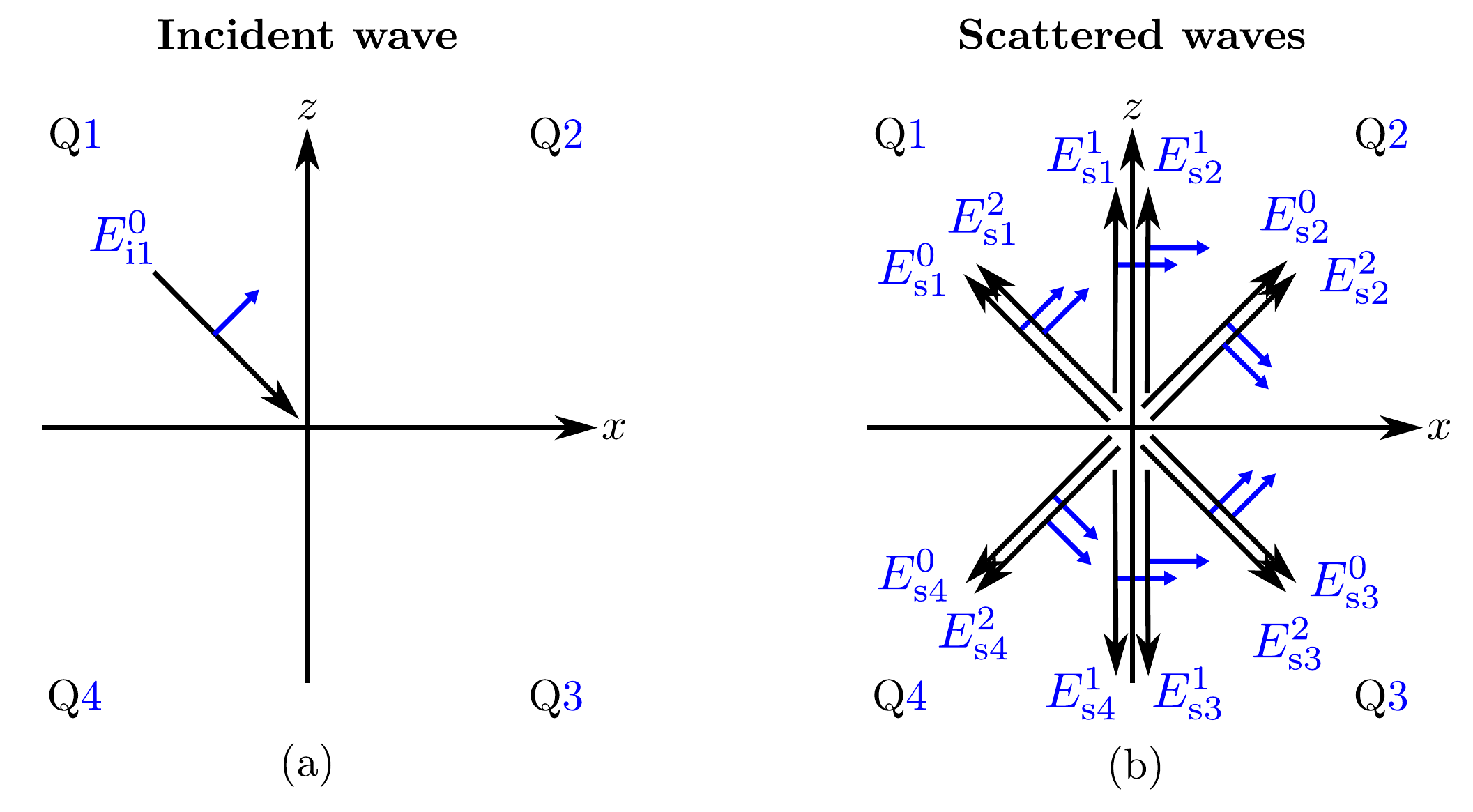}
\caption{Scattering situation with redundant diffraction channels. (a) A TM polarized wave is incident on a grating with the parameter set $\{1,1,45^\circ,\sqrt{2}/2\}$. (b) The grating diffracts the incident wave into diffraction orders 0, 1 and 2 leading to a redundant definition of the open channels as defined in Fig.~\ref{fig_do2chan}.}
\label{fig_redundantDO}
\end{figure}

\B{The first situation happens if $\hat{k}_{x,\text{i}}$ is an integer multiple of $d \mathcal{G}$ in~\eqref{eq_GEQ}. This implies that, for a given diffraction order $d$, its tangential wavenumber $\hat{k}_{x,\text{s}}^{d}$ may either be equal to $0$ or to $\hat{k}_{x,\text{i}}$. In both cases, applying the channel numbering convention defined in Fig.~\ref{fig_do2chan} leads a redundant definition of the open channels because different diffraction orders end-up overlapping each other at identical positions but in longitudinally opposite quadrants. Such a situation is depicted, for TM polarization only, in Fig.~\ref{fig_redundantDO}, which corresponds to a grating with the parameter set $\{1,1,45^\circ,\sqrt{2}/2\}$. \B{Since $\theta_\text{i}=45^\circ$ we have that $\hat{k}_{x,\text{i}}=\sqrt{2}/2$ which is indeed a multiple integer of the normalized grating period $\mathcal{G}=\sqrt{2}/2$.}}
In Fig.~\ref{fig_redundantDO}, the incident wave is impinging on the array from channel $0$ in Q1 and is scattered in all quadrants producing the diffraction orders 0, 1 and 2. Due to the chosen set of grating parameters, the diffraction orders 0 and 2 overlap themselves once the convention in Fig.~\ref{fig_do2chan} is applied, as can be seen in Fig.~\ref{fig_redundantDO}b. This means that the diffraction order channels 2 are redundant with respect to the $0^{\text{th}}$-order ones. To remove this redundancy, the total number of diffraction orders in a given region, computed using~\eqref{eq_Ntot}, should be reduced from $N_\text{a}$ to $N_\text{a} - (N_\text{a} - 1)/2$, which reduces to total size the global scattering matrix $\te{S}$. Moreover, we also need to force equality between the diffraction order channels $1$ that lie in longitudinally adjacent quadrants, i.e., $E_\text{s1}^{1}=E_\text{s2}^{1}$ and $E_\text{s3}^{1}=E_\text{s4}^{1}$ and so on. Generally, this is achieved by enforcing the following identities:
\B{
\begin{equation} 
    \label{eq_equality} 
\begin{split} 
    S_{q2}^{mn_0} = S_{q1}^{mn_0} &\quad\text{and}\quad S_{2q}^{m_0n} = S_{1q}^{m_0n},\\
    S_{q4}^{mn_0} = S_{q3}^{mn_0} &\quad\text{and}\quad S_{4q}^{m_0n} = S_{3q}^{m_0n},\\
    S_{q6}^{mn_0} = S_{q5}^{mn_0} &\quad\text{and}\quad S_{6q}^{m_0n} = S_{5q}^{m_0n},\\
    S_{q8}^{mn_0} = S_{q7}^{mn_0} &\quad\text{and}\quad S_{8q}^{m_0n} = S_{7q}^{m_0n},
\end{split}
\end{equation}
where $q=\{1 \dots 8\}$ and $m_{0}$ and $n_{0}$ are the diffraction order channels for which $\hat{k}_{x,\text{s}}= 0$.}

\B{The second situation occurs if $\hat{k}_{x,\text{s}}^{d_{1}} = - \hat{k}_{x,\text{s}}^{d_{2}}$ for any combination of the diffraction order numbers $\{d_1,d_2\}$ except $d_{1} = d_{2} = 0$. \B{Using~\eqref{eq_GEQ}, we may reformulate this condition as $\hat{k}_{x,\text{i}} = \left( d_{1}+d_{2} \right)\mathcal{G}/2$, which would, for instance, be problematic for a situation defined by the parameter set $D = \{1.5,1,45^\circ,\sqrt{2}/2\}$ and for the diffraction orders $d_{1} = 1$ and $d_{2} = 2$.} This situation is problematic because it also interferes with the channel numbering convention used in Fig.~\ref{fig_do2chan} leading to an overlap of different diffraction order channels, which is again an undesired redundancy. However, contrary to the first problematic situation discussed above, this case does not lead to any diffraction order with $\hat{k}_{x,\text{s}}=0$. In this case, to remove the redundancy, the total number of diffraction orders computed with~\eqref{eq_Ntot} should be reduced from $N_\text{a}$ to $N_\text{a}/2$.}

\section{Relation to spatial symmetries}
\label{sec_sym} 

We are now interested in investigating the scattering symmetries of the global scattering matrix $\te{S}$ and its sub-scattering matrices $\te{S}^{mn}$. For this purpose, we follow an approach similar to the one described in~\cite{dmitriev2011,dmitriev2013a,achouri2023a}, which consists in defining matrix representations of various symmetry operations. Concretely, we want to find a matrix $\te{M}_{\Lambda}$, where $\Lambda$ corresponds to a given symmetry operation, that transforms a field vector $\ve{E}^{d}_\text{a}$ in Fig.~\ref{fig_schematicdo} into a new field vector $\ve{E}^{'d}_\text{a}$ that corresponds to the application of the chosen symmetry operation. This may be written as
\begin{equation} 
    \label{eq_fieldbasischange} 
 \ve{E}^{'d}_\text{a} = \te{M}_{\Lambda}\cdot\ve{E}^{d}_\text{a},
\end{equation}
where $a=\left\{ \text{i}, \text{s} \right\}$.

We shall next restrict our attention to symmetry operations that are consistent with a diffraction grating producing diffraction orders in the $xz$-plane only. We thus consider the following set of seven different symmetry operations: $\sigma_{x}$, $\sigma_{y}$, $\sigma_{z}$, $C_{2x}$, $C_{2y}$, $C_{2z}$ and $i$, which respectively correspond to mirror symmetries, $180^\circ$-rotation symmetries and inversion symmetry.

\begin{figure}[h!]
\centering
\includegraphics[width=\linewidth]{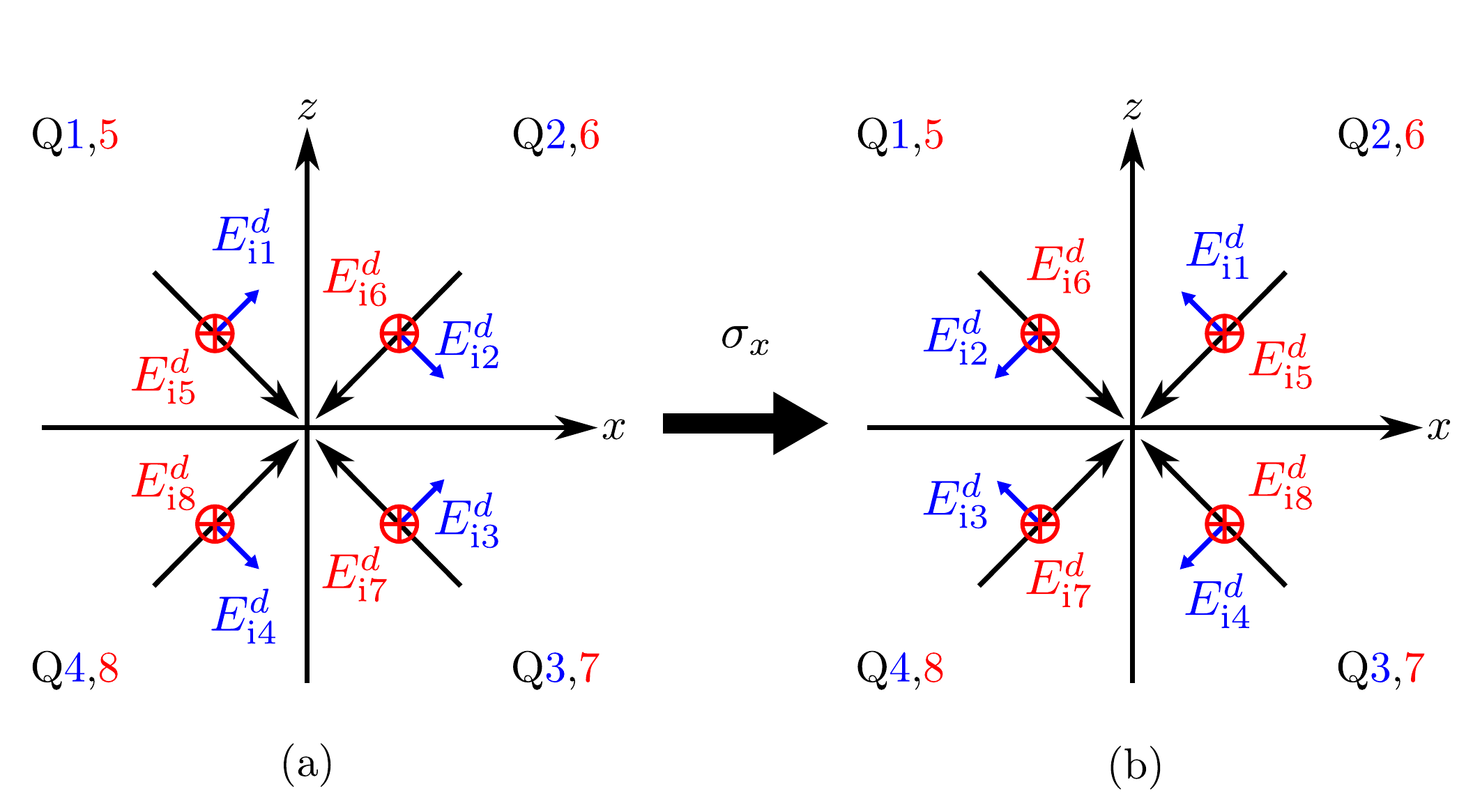}
    \caption{Application of the symmetry operation $\sigma_{x}$ on the system depicted in Fig.~\ref{fig_schematicdo}a. (a) Initial situation identical to the one in Fig.~\ref{fig_schematicdo}a. (b) Situation after the application of $\sigma_{x}$.}
\label{fig_schematicdosigx}
\end{figure}

As an example, we now illustrate how to obtain the matrix representations $\te{M}_{\sigma_{x}}$ that corresponds to the symmetry operation $\sigma_{x}$ whose effect on the system in Fig.~\ref{fig_schematicdo}a is depicted in Fig.~\ref{fig_schematicdosigx}.
As can be seen, the operation $\sigma_{x}$ does not induce any rotation of polarization. However, it moves the waves that are initially in quadrants Q1 (Q5) into Q2 (Q6) and those in Q4 (Q8) into Q3 (Q7), and vice versa. Additionally, TE and TM waves are phase shifted by 0 and $\pi$, respectively. Applying the same analysis to all the other symmetry operations, we find that the corresponding representation matrices are given by
\begin{equation} 
\begin{split} 
\te{M}_{\sigma_x} &= 
\begin{bmatrix}
-\te{A} & \te{0} \\
\te{0}&   \te{A}
\end{bmatrix},\quad\,\,\,
\te{M}_{\sigma_y} = 
\begin{bmatrix}
\te{I} & \te{0} \\
\te{0}&   -\te{I}
\end{bmatrix},\\
\te{M}_{\sigma_z} &= 
\begin{bmatrix}
\te{B} & \te{0} \\
\te{0}&   \te{B}
\end{bmatrix},\quad\,\,\,\,
\te{M}_{C_{2x}} = 
\begin{bmatrix}
\te{B} & \te{0} \\
\te{0}&   -\te{B}
\end{bmatrix},\\
\te{M}_{C_{2y}} &= 
\begin{bmatrix}
-\te{C} & \te{0} \\
\te{0}&   \te{C}
\end{bmatrix},\quad
\te{M}_{C_{2z}} = 
\begin{bmatrix}
-\te{A} & \te{0} \\
\te{0}&   -\te{A}
\end{bmatrix},\\
\te{M}_{i} &= 
\begin{bmatrix}
-\te{C} & \te{0} \\
\te{0}&   -\te{C}
\end{bmatrix},
\end{split}
\end{equation}
where $\te{0}$ is a $4\times 4$ zero matrix and the sub-matrices $\te{A}, \te{B}$ and $\te{C}$ are given by
\begin{equation}
\label{eq_ABC} 
\begin{split}
\te{A} = &
\begin{bmatrix}
0 & 1 & 0 & 0 \\
1 & 0 & 0 & 0 \\
0 & 0 & 0 & 1 \\
0 & 0 & 1 & 0 
\end{bmatrix},\quad
\te{B} = 
\begin{bmatrix}
0 & 0 & 0 & 1 \\
0 & 0 & 1 & 0 \\
0 & 1 & 0 & 0 \\
1 & 0 & 0 & 0 
\end{bmatrix},\\
&\quad\quad\quad\te{C} = 
\begin{bmatrix}
0 & 0 & 1 & 0 \\
0 & 0 & 0 & 1 \\
1 & 0 & 0 & 0 \\
0 & 1 & 0 & 0 
\end{bmatrix}.
\end{split}
\end{equation}

Now that we are able to describe the effects that these symmetry operations have on the waves interacting with the grating, we derive the invariance condition that applies to the grating scattering matrix. Since the matrix $\te{M}_{\Lambda}$ applies to an amplitude vector $\ve{E}_\text{a}^{d}$, we combine~\eqref{eq_fieldbasischange} with~\eqref{eq_subS} to obtain the transformed sub-scattering matrix $\te{S}^{'mn}$ that represents the scattering response of the grating between the channels $m$ and $n$ after having applied a given symmetry operation $\Lambda$. The transformation from $\te{S}^{mn}$ to $\te{S}^{'mn}$ corresponds to a change of basis and can be easily derived by combining~\eqref{eq_subS} and~\eqref{eq_fieldbasischange} leading to
\begin{equation}
\te{S}^{'mn} = \te{M}_\Lambda\cdot\te{S}^{mn}\cdot\te{M}^{-1}_\Lambda.
\end{equation}
If the grating is invariant under a given symmetry operation, then we must have that $\te{S}^{'mn}=\te{S}^{mn}$, which leads to the invariance condition
\begin{equation}
\label{eq_inv} 
\te{S}^{mn} = \te{M}_\Lambda\cdot\te{S}^{mn}\cdot\te{M}^{-1}_\Lambda.
\end{equation}
Equation~\eqref{eq_inv} represents the main result of this paper. It corresponds to a system of equations that may be solved following a procedure similar to the one described in~\cite{achouri2023a}. This procedure consists in starting with a full $\te{S}^{mn}$ matrix. Then, for each symmetries that the grating exhibits, we solve~\eqref{eq_inv} with the corresponding $\te{M}_{\Lambda}$, which typically sets to zero some components in $\te{S}^{mn}$ or forces some other components to be equal to each other. We then repeat this process for all of the grating symmetries, which ultimately gives us the final reduced form of $\te{S}^{mn}$. Finally, we apply this process to all possible combinations of $m$ and $n$ to obtain the final form of the grating global scattering matrix $\te{S}$.

Note that in addition to these properties of symmetry, the global scattering matrix $\te{S}$ is reciprocal if it satisfies the reciprocity condition $\te{S} = \te{S}^T$, which implies that the corresponding sub-scattering matrices must satisfy
\begin{equation}
\te{S}^{mm} = \left( \te{S}^{mm} \right)^T \quad \text{and} \quad \te{S}^{mn} = \left( \te{S}^{nm} \right)^T,
\end{equation}
where the superscript $T$ denotes the transpose operation.

Finally, note that in order to facilitate the use of the proposed technique, we have created a Python utility that automatically computes the diffraction orders, the scattering symmetries and the global scattering matrix that correspond to a given diffraction grating~\cite{gratingsym2026}.


\section{Illustrative examples}

\begin{figure*}[t!]
\centering
\includegraphics[width=\linewidth]{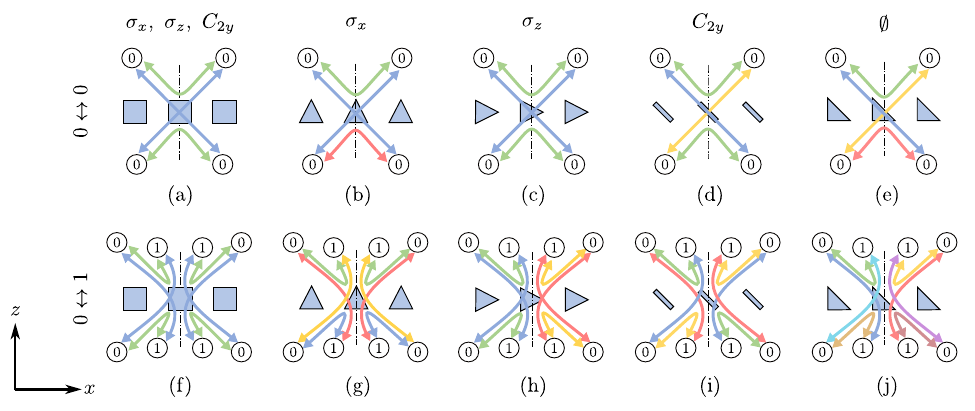}
\caption{Scattering symmetries for various gratings. The gratings are assumed to either be infinite along the $y$-direction or at least be $\sigma_{y}$ symmetric. Top row: scattering between 0$^{\text{th}}$-order channels. Bottom row: scattering between 0$^{\text{th}}$-order and 1$^{\text{st}}$-order channels. Identical arrow colors indicate identical scattering coefficients.}
\label{fig_symorders}
\end{figure*}

Let us now illustrate the application of the proposed methodology. For this purpose, we start by considering the reciprocal diffraction grating of Fig.~\ref{fig_do2chan} that corresponds to the parameter set $D_{1}$ discussed previously. For simplicity, we consider that this grating is either infinite along the $y$-direction or is at least $\sigma_{y}$ symmetric meaning that the grating is unable to scatter cross-polarized waves. We now investigate the scattering symmetries of this grating by considering different possible geometries for its scattering particles. Specifically, we consider 5 different geometries corresponding to arrays of squares ($\sigma_{x}, \sigma_{z}, C_{2y}$), vertical triangles ($\sigma_{x}$), horizontal triangles ($\sigma_{z}$), tilted rods ($C_{2y}$) and rotated triangles (exhibiting no symmetries with respect to the $x$, $y$ and $z$ directions). These different gratings are depicted in Fig.~\ref{fig_symorders} along with the corresponding scattering symmetries between both $0^{\text{th}}$ and $1^{\text{st}}$ order channels. To compute these scattering symmetries, we use the formalism described in Secs.~\ref{sec_theo} and~\ref{sec_sym}. To illustrate this process, we now apply it to the first grating in Fig.~\ref{fig_symorders} that corresponds to an array of squares. 

With the parameter set $D_{1}$, the only two open diffraction order channels on both sides of the grating are the channels 0 and 1. From the theory in Sec.~\ref{sec_theo}, this means that the grating global scattering matrix is
\begin{equation} 
\te{S} = \begin{bmatrix} 
    \te{S}^{00} & \te{S}^{01} \\
    \te{S}^{10} & \te{S}^{11}
\end{bmatrix}.  
\end{equation}
By reciprocity, we know that $\te{S} = \te{S}^T$ implying that
\begin{equation} 
    \label{eq_recip2} 
\te{S}^{01} = \left( \te{S}^{10} \right)^T,\ \te{S}^{00} = \left( \te{S}^{00} \right)^T,\ \te{S}^{11} = \left( \te{S}^{11} \right)^T.
\end{equation}
We now apply the theory discussed in Sec.~\ref{sec_sym} to obtain the shape of the sub-scattering matrices $\te{S}^{00}$ and $\te{S}^{01}$. It follows that $\te{S}^{00}$ reads
\small{
\begin{equation} 
\label{eq_S00} 
\te{S}^{00}=
\begin{bmatrix}
0 & R_\text{pp}^{00} & T_\text{pp}^{00} & 0 & 0 & 0 & 0 & 0\\
R_\text{pp}^{00} & 0 & 0 & T_\text{pp}^{00} & 0 & 0 & 0 & 0\\
T_\text{pp}^{00} & 0 & 0 & R_\text{pp}^{00} & 0 & 0 & 0 & 0\\
0 & T_\text{pp}^{00} & R_\text{pp}^{00} & 0 & 0 & 0 & 0 & 0\\
0 & 0 & 0 & 0 & 0 & R_\text{ss}^{00} & T_\text{ss}^{00} & 0\\
0 & 0 & 0 & 0 & R_\text{ss}^{00} & 0 & 0 & T_\text{ss}^{00}\\
0 & 0 & 0 & 0 & T_\text{ss}^{00} & 0 & 0 & R_\text{ss}^{00}\\
0 & 0 & 0 & 0 & 0 & T_\text{ss}^{00} & R_\text{ss}^{00} & 0
\end{bmatrix},
\end{equation}
}
where $R$ and $T$ denote arbitrary reflection and transmission coefficients and the subscripts ``p'' and ``s'' refer to parallel (TM) and perpendicular (TE) polarizations. We see that $\te{S}^{00}$ corresponds to a reduced form of the forward scattering matrix in~\eqref{eq_FS}. This is because the application of the symmetry operation $\sigma_{y}$ forces all cross-polarized coefficients (top-right and bottom-left $4\times 4$ elements) to be zero, whereas the application of the operations $\sigma_{x}$ and $\sigma_{z}$ ($C_{2y}$ being redundant here) along with reciprocity forces the matrix to be symmetric, i.e., equal to its transpose. From~\eqref{eq_S00}, we see that the reflection coefficients between quadrants Q1 (Q5) and Q2 (Q6) and between Q3 (Q7) and Q4 (Q8) must be identical, as are the transmission coefficients between quadrants Q1 (Q5) and Q3 (Q7) and between Q2 (Q6) and Q4 (Q8), which is precisely how it is depicted in Fig.~\ref{fig_symorders}a, where colored arrows are used to depict these scattering parameters.

By the same token, the matrix $\te{S}^{10}$ reads
\small{
\begin{equation} 
\label{eq_S01} 
\te{S}^{10}=
\begin{bmatrix}
R_\text{pp}^{10} & 0 & 0 & T_\text{pp}^{10} & 0 & 0 & 0 & 0 \\
0 & R_\text{pp}^{10} & T_\text{pp}^{10} & 0 & 0 & 0 & 0 & 0 \\
0 & T_\text{pp}^{10} & R_\text{pp}^{10} & 0 & 0 & 0 & 0 & 0 \\
T_\text{pp}^{10} & 0 & 0 & R_\text{pp}^{10} & 0 & 0 & 0 & 0 \\
0 & 0 & 0 & 0 & R_\text{ss}^{10} & 0 & 0 & T_\text{ss}^{10} \\
0 & 0 & 0 & 0 & 0 & R_\text{ss}^{10} & T_\text{ss}^{10} & 0 \\
0 & 0 & 0 & 0 & 0 & T_\text{ss}^{10} & R_\text{ss}^{10} & 0 \\
0 & 0 & 0 & 0 & T_\text{ss}^{10} & 0 & 0 & R_\text{ss}^{10}
\end{bmatrix}.
\end{equation}
}
Again, all reflection and transmission coefficients between the $0^{\text{th}}$- and $1^{\text{st}}$-order channels must be equal due to both reciprocity and symmetry, as depicted in Fig.~\ref{fig_symorders}f. Finally, the matrix $\te{S}^{01}$ is found using~\eqref{eq_recip2}, whereas the matrix $\te{S}^{11}$ has a shape identical to that of $\te{S}^{00}$ given in~\eqref{eq_S00} and is therefore not presented here.

The scattering analysis of the four other gratings follows exactly the same procedure and is therefore omitted for conciseness. Only the corresponding scattering symmetries are plotted in Fig.~\ref{fig_symorders}. Notice, how, for each case, the diagrams are perfectly consistent with the prescription of reciprocity and the spatial symmetries of the grating.\\

\B{We now consider an application of this theory to the following design problem: how to implement a reciprocal and passive device that exhibits angle-asymmetric transmission in intensity, i.e., $|T(\theta)|^2 \neq |T(-\theta)|^2$. Specifically, we want to achieve $|T(\theta)|^2 = 1$ and $|T(-\theta)|^2 =0$, as discussed in~\cite{luo2021,fan2021,abouelatta2025,shaham2026}.
Intuitively, one might think that, since the waves propagate in the $xz$-plane, to break angular transmission symmetry, it would be sufficient to break the grating mirror symmetry $\sigma_{x}$. However, referring to Fig.~\ref{fig_symorders}c, we see that this does not work. The reason is that the grating in Fig.~\ref{fig_symorders}c is $\sigma_{z}$ symmetric and \emph{reciprocal}, which forces all transmission coefficients to be equal to each other as well as all reflection coefficients. An actual solution to implement the desired design is to consider a grating exhibiting broken $\sigma_{x}$ and $\sigma_{z}$ symmetries, like the one in Fig.~\ref{fig_symorders}e, which is the design approach proposed in~\cite{shaham2026} for the microwave regime. Another simpler solution in terms of nanofabrication for the optical regime consists in implementing a grating with $C_{2y}$ symmetry as the one in Fig.~\ref{fig_symorders}d. An example of such a design is shown in~\cite{abouelatta2025}, where a one-dimensional diffraction grating with $C_{2y}$ symmetry is implemented to realize a step function in momentum space.}

\B{Note that the metagrating in~\cite{abouelatta2025} achieves angular transmission asymmetry in a \emph{lossless} fashion by redirecting the energy, for a $-\theta$ incidence angle, into the $+1$ diffraction order in reflection, effectively achieving $|T(-\theta)|^2 =0$, while allowing the energy to transmit, for a $+\theta$ incidence angle, into the $0^\text{th}$-diffraction order in transmission leading to $|T(\theta)|^2 =1$. This corresponds to exploiting both the scattering channels of Fig.~\ref{fig_symorders}d and Fig.~\ref{fig_symorders}i.}

\B{Another possibility to achieve angular transmission asymmetry is the one presented in~\cite{luo2021,fan2021}, where a subwavelength (diffractionless) metasurface is designed with \emph{lossy} scattering particles, as in Fig.~\ref{fig_symorders}d. In this case, the metasurface achieves the desired asymmetric scattering response via asymmetric absorption rather than the exploitation of diffraction channels, as in~\cite{abouelatta2025}.}

\section{Extrinsic Chirality in Gratings}

Let us now again consider the first grating in Fig.~\ref{fig_symorders} and assume that it is \emph{asymmetric} in the $y$-direction, i.e., broken $\sigma_{y}$ symmetry. Applying again the formalism described above yields the following $\te{S}^{00}$ and $\te{S}^{10}$ matrices:
\small{
\begin{equation} 
    \label{eq_S00ps} 
\te{S}^{00}=
\begin{bmatrix}
0 & R_\text{pp}^{00} & T_\text{pp}^{00} & 0 & 0 & -R_\text{ps}^{00} & -T_\text{ps}^{00} & 0\\
R_\text{pp}^{00} & 0 & 0 & T_\text{pp}^{00} & R_\text{ps}^{00} & 0 & 0 & T_\text{ps}^{00}\\
T_\text{pp}^{00} & 0 & 0 & R_\text{pp}^{00} & T_\text{ps}^{00} & 0 & 0 & R_\text{ps}^{00}\\
0 & T_\text{pp}^{00} & R_\text{pp}^{00} & 0 & 0 & -T_\text{ps}^{00} & -R_\text{ps}^{00} & 0\\
0 & R_\text{ps}^{00} & T_\text{ps}^{00} & 0 & 0 & R_\text{ss}^{00} & T_\text{ss}^{00} & 0\\
-R_\text{ps}^{00} & 0 & 0 & -T_\text{ps}^{00}   & R_\text{ss}^{00} & 0 & 0 & T_\text{ss}^{00}\\
-T_\text{ps}^{00} & 0 & 0 & -R_\text{ps}^{00}   & T_\text{ss}^{00} & 0 & 0 & R_\text{ss}^{00}\\
0 & T_\text{ps}^{00} & R_\text{ps}^{00} & 0 & 0 & T_\text{ss}^{00} & R_\text{ss}^{00} & 0
\end{bmatrix},
\end{equation}
}
and
\small{
\begin{equation} 
    \label{eq_S01ps} 
\te{S}^{10}=
\begin{bmatrix}
R_\text{pp}^{10} & 0 & 0 & T_\text{pp}^{10} & R_\text{ps}^{10} & 0 & 0 & T_\text{ps}^{10} \\
0 & R_\text{pp}^{10} & T_\text{pp}^{10} & 0 & 0 & -R_\text{ps}^{10} & -T_\text{ps}^{10} & 0 \\
0 & T_\text{pp}^{10} & R_\text{pp}^{10} & 0 & 0 & -T_\text{ps}^{10} & -R_\text{ps}^{10} & 0 \\
T_\text{pp}^{10} & 0 & 0 & R_\text{pp}^{10} & T_\text{ps}^{10} & 0 & 0 & R_\text{ps}^{10} \\
R_\text{sp}^{10} & 0 & 0 & T_\text{sp}^{10} & R_\text{ss}^{10} & 0 & 0 & T_\text{ss}^{10} \\
0 & -R_\text{sp}^{10} & -T_\text{sp}^{10} & 0 & 0 & R_\text{ss}^{10} & T_\text{ss}^{10} & 0 \\
0 & -T_\text{sp}^{10} & -R_\text{sp}^{10} & 0 & 0 & T_\text{ss}^{10} & R_\text{ss}^{10} & 0 \\
T_\text{sp}^{10} & 0 & 0 & R_\text{sp}^{10} & T_\text{ss}^{10} & 0 & 0 & R_\text{ss}^{10}
\end{bmatrix}.
\end{equation}
}
In both scattering matrices, we notice that the top-left and bottom-right $4\times 4$ elements, corresponding to co-polarized scattering parameters, remain identical to those in~\eqref{eq_S00} and~\eqref{eq_S01}. We also notice the appearance of cross-polarized scattering parameters, some of them being equal to minus each other due to the involved symmetries. These cross-polarized scattering parameters correspond to extrinsic chiral effects that stem from the fact that this grating is simultaneously illuminated in the $xz$-plane while being asymmetric along the $y$-direction~\cite{plum2009a,kim2021,achouri2023a}. The combined consideration of illumination direction and broken spatial symmetries leads to a system (illumination + physical structure) with all mirror symmetries being broken that is consequently able to generate effective chiral responses.

\begin{figure*}[t!]
\centering
\includegraphics[width=0.72\linewidth]{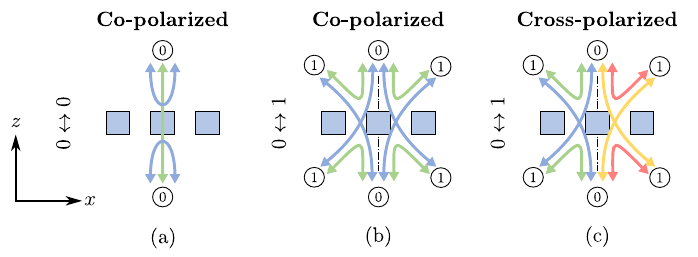}
    \caption{Grating exhibiting extrinsic chiral responses when illuminated at normal incidence. The grating is described by the parameter set $\{1,1,0,\sqrt{2}/2\}$ and thus produces the diffraction order channels 0 and 1. (a) Only co-polarized scattering is possible between the $0^{\text{th}}$-order channels. (b) Co-polarized and (c) cross-polarized scattering between the $0^{\text{th}}$- and $1^{\text{st}}$-order channels. Identical arrow colors indicate identical scattering coefficients.}
\label{fig_chiral}
\end{figure*}

Interestingly, in~\eqref{eq_S00ps}, the top-right and bottom-left $4\times 4$ elements are simply the transpose of each other due to reciprocity. This means that, for instance, the TE-to-TM transmission coefficient from Q1 to Q7 must be equal to the TM-to-TE transmission coefficient from Q7 to Q1, which is exactly what we expect from reciprocity. However, in~\eqref{eq_S01ps}, we see that the corresponding $4\times 4$ elements are not equal to each other. This shows that, while the system remains fully reciprocal, the cross-polarized scattering between different diffraction orders ($0 \leftrightarrow 1$) is not restricted in the same way that it is between identical diffraction orders ($0 \leftrightarrow 0$ or $1 \leftrightarrow 1$). This demonstrates that diffraction gratings offer more degrees of freedom for controlling the propagation of electromagnetic waves compared to diffraction-less systems like, for instance, metasurfaces.

We have just considered the case of extrinsic chirality arising in a diffraction grating illuminated at oblique incidence. While interesting, the emergence of extrinsic chiral effects in such a configuration is expected, since these effects are typically observed under oblique illumination~\cite{plum2009a,kim2021,achouri2023a}. A more intriguing question is therefore whether a diffraction grating illuminated at \emph{normal} incidence can also give rise to extrinsic chiral effects.

To investigate this case, we now consider a diffraction grating described by the parameter set $\{1,1,0,\sqrt{2}/2\}$. Since the grating period is slightly larger than the wavelength, it generates the $0^{\text{th}}$ and $1^{\text{st}}$ diffraction orders in both $\pm z$ regions. We also consider that the grating is $\sigma_{x}$ and $\sigma_{z}$ symmetric and $\sigma_{y}$ asymmetric. Since some of the waves that interact with this grating propagate normally with respect to it, we need to carefully apply the corrections discussed in Fig.~\ref{fig_redundantDO}, to account for the presence of redundant diffraction order channels, and properly equate the corresponding scattering coefficients using~\eqref{eq_equality}.

We now apply the formalism described above and depict the corresponding co- and cross-polarized scattering responses of this grating in Fig.~\ref{fig_chiral}. Note that the different colors between both longitudinal sides in Fig.~\ref{fig_chiral}c are due to the chosen convention for the TE and TM polarizations in Fig.~\ref{fig_schematicdo}. Considering only an illumination coming at normal incidence, i.e., from the diffraction channel number 0, the corresponding sub-scattering matrices are given by
\small{
\begin{equation} 
\label{eq_S00chiral} 
\te{S}^{00}=
\begin{bmatrix}
R_\text{pp}^{00} & R_\text{pp}^{00} & T_\text{pp}^{00} & T_\text{pp}^{00} & 0 & 0 & 0 & 0\\
R_\text{pp}^{00} & R_\text{pp}^{00} & T_\text{pp}^{00} & T_\text{pp}^{00} & 0 & 0 & 0 & 0\\
T_\text{pp}^{00} & T_\text{pp}^{00} & R_\text{pp}^{00} & R_\text{pp}^{00} & 0 & 0 & 0 & 0\\
T_\text{pp}^{00} & T_\text{pp}^{00} & R_\text{pp}^{00} & R_\text{pp}^{00} & 0 & 0 & 0 & 0\\
0 & 0 & 0 & 0 & R_\text{ss}^{00} & R_\text{ss}^{00} & T_\text{ss}^{00} & T_\text{ss}^{00}\\
0 & 0 & 0 & 0 & R_\text{ss}^{00} & R_\text{ss}^{00} & T_\text{ss}^{00} & T_\text{ss}^{00}\\
0 & 0 & 0 & 0 & T_\text{ss}^{00} & T_\text{ss}^{00} & R_\text{ss}^{00} & R_\text{ss}^{00}\\
0 & 0 & 0 & 0 & T_\text{ss}^{00} & T_\text{ss}^{00} & R_\text{ss}^{00} & R_\text{ss}^{00}
\end{bmatrix},
\end{equation}
}
and
\small{
\begin{equation} 
\label{eq_S01chiral} 
\te{S}^{10}=
\begin{bmatrix}
R_\text{pp}^{10}  & R_\text{pp}^{10}  & T_\text{pp}^{10}  & T_\text{pp}^{10}  & R_\text{ps}^{10}  & R_\text{ps}^{10}  & -T_\text{ps}^{10} & -T_\text{ps}^{10}\\
R_\text{pp}^{10}  & R_\text{pp}^{10}  & T_\text{pp}^{10}  & T_\text{pp}^{10}  & -R_\text{ps}^{10} & -R_\text{ps}^{10} & T_\text{ps}^{10}  & T_\text{ps}^{10}\\
T_\text{pp}^{10}  & T_\text{pp}^{10}  & R_\text{pp}^{10}  & R_\text{pp}^{10}  & T_\text{ps}^{10}  & T_\text{ps}^{10}  & -R_\text{ps}^{10} & -R_\text{ps}^{10}\\
T_\text{pp}^{10}  & T_\text{pp}^{10}  & R_\text{pp}^{10}  & R_\text{pp}^{10}  & -T_\text{ps}^{10} & -T_\text{ps}^{10} & R_\text{ps}^{10}  & R_\text{ps}^{10}\\
R_\text{sp}^{10}  & R_\text{sp}^{10}  & -T_\text{sp}^{10} & -T_\text{sp}^{10} & R_\text{ss}^{10}  & R_\text{ss}^{10}  & T_\text{ss}^{10}  & T_\text{ss}^{10}\\
-R_\text{sp}^{10} & -R_\text{sp}^{10} & T_\text{sp}^{10}  & T_\text{sp}^{10}  & R_\text{ss}^{10}  & R_\text{ss}^{10}  & T_\text{ss}^{10}  & T_\text{ss}^{10}\\
T_\text{sp}^{10}  & T_\text{sp}^{10}  & -R_\text{sp}^{10} & -R_\text{sp}^{10} & T_\text{ss}^{10}  & T_\text{ss}^{10}  & R_\text{ss}^{10}  & R_\text{ss}^{10}\\
-T_\text{sp}^{10} & -T_\text{sp}^{10} & R_\text{sp}^{10}  & R_\text{sp}^{10}  & T_\text{ss}^{10}  & T_\text{ss}^{10}  & R_\text{ss}^{10}  & R_\text{ss}^{10}
\end{bmatrix}.
\end{equation}
}
As expected, the scattering between the $0^{\text{th}}$-order diffraction channels, which corresponds to waves that propagate normally with respect to the grating, do not exhibit any cross-polarization effects. However, the scattering between $0^{\text{th}}$- and $1^{\text{st}}$-order diffraction channels exhibits both co- and cross-polarized components. Examining~\eqref{eq_S01chiral}, we see that this type of scattering does indeed correspond to extrinsic chirality. The reason for this is that, despite the fact that the incident wave impinges normally on the grating, the $1^{\text{st}}$ diffraction orders propagate obliquely, which contributes to breaking all the system mirror symmetries. 

\section{Conclusion}

We have developed a formalism to systematically compute the scattering symmetries of a diffraction grating. Based on this formalism, we have investigated the scattering symmetries of several gratings exhibiting various spatial symmetries. We have seen that it is not always intuitive to predict the scattering symmetries of systems having several open diffraction order channels, especially when reciprocity is involved. This highlights the usefulness of the proposed approach.

The proposed method relies on the construction of a global scattering matrix that connects all possible incident and scattered diffraction channels. By introducing matrix representations of the relevant spatial symmetry operations, we derived an invariance condition that directly constrains the elements of the sub-scattering matrices associated with each pair of diffraction orders. This approach makes it possible to straightforwardly determine which scattering coefficients must vanish and which must be equal to each other for a given set of structural symmetries and reciprocity conditions.

Using this framework, we examined several diffraction gratings with various spatial symmetries and showed that they lead to distinct scattering symmetries. We also illustrated how the formalism can help to design devices with engineered diffraction responses, such as reciprocal structures exhibiting angle-asymmetric transmission. In addition, we analyzed the emergence of extrinsic chiral effects in diffraction gratings. In particular, we showed that such effects can arise not only under oblique illumination but also, and more interestingly, under normal incidence but only when higher diffraction orders are present and propagate obliquely and the grating lacks certain spatial symmetries.

Note that, while the present work focused on configurations where diffraction occurs within the plane of incidence, the approach can be extended to more general situations involving fully three-dimensional diffraction and arbitrary lattice geometries.

\begin{acknowledgments}
We acknowledge funding from the Swiss National Science Foundation (project TMSGI2\_218392).
\end{acknowledgments}

\bibliographystyle{ieeetr}
\bibliography{../../../../References/All.bib}
	
\end{document}